\documentclass[sigconf]{acmart}

\usepackage{booktabs} 

\begin{document}
\title{Evaluating Singleplayer and Multiplayer in Human Computation Games}

\author{Kristin Siu}
\affiliation{\institution{Georgia Institute of Technology}}
\email{kasiu@gatech.edu}

\author{Matthew Guzdial}
\affiliation{\institution{Georgia Institute of Technology}}
\email{mguzdial3@gatech.edu}

\author{Mark O. Riedl}
\affiliation{\institution{Georgia Institute of Technology}}
\email{riedl@cc.gatech.edu}

\renewcommand{\shortauthors}{Siu et al.}

\begin{abstract}
Human computation games (HCGs) can provide novel solutions to intractable computational problems, help enable scientific breakthroughs, and provide datasets for artificial intelligence.
However, our knowledge about how to design and deploy HCGs that appeal to players and solve problems effectively is incomplete.
We present an investigatory HCG based on \emph{Super Mario Bros.} 
We used this game in a human subjects study to investigate how different social conditions---singleplayer and multiplayer---and scoring mechanics---collaborative and competitive---affect players' subjective experiences, accuracy at the task, and the completion rate.
In doing so, we demonstrate a novel design approach for HCGs, and discuss the benefits and tradeoffs of these mechanics in HCG design.
\end{abstract}

%
%




\maketitle

\section{Introduction}
Human computation games (HCGs) have been used to tackle various computationally-intractable problems, such as classifying information and discovering scientific solutions, by leveraging the skills of human players.
These games, also known as Games with a Purpose, scientific discovery games, or citizen science games, ask players to complete crowdsourcing problems or \emph{tasks} by interacting with the mechanics of a game.
The successful solutions to problems such as image labeling, protein folding, and more have helped to demonstrate how games can be considered an effective alternative to traditional crowdsourcing platforms while providing potentially-entertaining experiences for their participants.

However, the process of developing a human computation game is often daunting.
Choosing to build a game requires significant time and investment, especially for scientists and task providers who may not have experience designing and developing games.
Complicating this is the fact that, unlike other games designed primarily for entertainment, HCGs serve dual purposes.
First, they must provide an engaging player experience.
Second, they must solve the corresponding human computation task effectively.
These two design goals reflect the preferences of players and the needs of human computation task providers (e.g., scientists or researchers) respectively, and ideally, HCGs afford both.
However, optimizing only for an engaging player experience may result in a game that does not properly or effectively solve the underlying task.
Meanwhile, optimizing only for solving the underlying task may result in an uninteresting game that will not attract or retain players, thus yielding few or poor task results.

When executed poorly, these games garner a reputation for being unengaging for players and ineffective for task providers \cite{tuite2014:gwap-problem}.
This is in spite of evidence that HCGs are an effective interface for crowdsourced work, even when compared directly with other crowdsourcing platforms \cite{krause2015:hcgs-vs-crowdsourcing}.
Understanding how gameplay mechanics affect both the player experience and completion of the human computation task is imperative in order to create effective games.
However, current HCG design is generally limited to templates and anecdotal examples of successful games, lacking generalized design knowledge or guidelines about how to develop games for new kinds of tasks and changing player audiences.


This means exploring the design space of HCGs and understanding how the elements of these games affect both the player experience and completion of the underlying task.
Specifically, we want to understand and focus on game \emph{mechanics}, the rules that dictate what interactions players can have with the game, as these are directly related to both the player experience and the process of solving the human computation task.

In this paper, we explore the question of how human computation games could benefit from co-located multiplayer game mechanics through a study comparing mechanical variations.
Many mechanics in HCGs are designed to facilitate consensus and agreement as a way to verify the results (e.g., two players providing answers to the same problem and getting rewarded with in-game points if they agree).
Traditionally, HCG players are isolated (i.e., prohibited from real-time communication during play) to prevent collusion that may impact results.
On the other hand, co-located multiplayer experiences have been shown to be strongly-engaging for players \cite{wehbe2015:colocated-play} and also in domains such as education (where games are intended to both teach and engage) \cite{ke2007:gameplaying-for-maths, plass2013:impact-individual-collab-compete-math-games}.
Ultimately, we want to understand if co-located gameplay mechanics, based on their success in other dual-purpose domains, might generalize to HCGs while verifying if, and how, collusion may negatively impact human computation task results.

To better examine HCG design questions, we built an HCG based on the classic game \emph{Super Mario Bros.}
Using this game, we ran a user study to compare singleplayer and co-located (i.e., local) multiplayer gameplay experiences.
Within the multiplayer experience, we also compare two variants: one using a collaborative scoring system and one using a competitive scoring system.

We evaluate these variations in gameplay mechanics using a human computation task with a known solution to see how changes in mechanics affect the aspects of the \emph{player experience} and the results of the human computation task (which we refer to as \emph{task completion}).
We also report on the results of a survey asking experts in HCGs for their opinions and perceptions of these specific mechanics.
Using our study results and reported expert opinions, we highlight and discuss four design implications of our work---adaptation of successful gameplay mechanics to HCGs, use of direct player communication, synchronous competitive play, and synchronous collaborative play---and how these impact the player experience and the completion of the human computation task.

The remainder of this paper is broken down as follows.
First, we review relevant work in the space of HCGs, game design, and studies of player behavior.
Next, we describe our \emph{Super Mario Bros.}-inspired HCG, \emph{Gwario}, and the methodology of our study.
We then present our results of the study, and report on expert opinions on relevant HCG mechanics.
This is followed by discussion of our design implications, contextualized by our results and expert opinions.
We conclude with limitations and future work.
\section{Related Work}
\subsection{Overview of HCGs}
\label{sec:hcg-overview}
Human computation games have been used to solve a wide variety of tasks, often those which are considered computationally-intractable for current algorithms and/or require commonsense human knowledge or reasoning.
These games are also known as Games with a Purpose (GWAPs), scientific discovery games, citizen science games, and crowdsourcing games.
The earliest examples of HCGs asked players to act as human classifiers to annotate or label data, such as images \cite{vonahn2004:esp}, music \cite{barrington2012:herdit}, relational information \cite{carranza2012:ontogalaxy-vs-esp}, and galaxy clustering \cite{lintott2008:galaxyzoo}.
Other HCGs utilize players to assist with scientific optimization problems, such as protein folding \cite{cooper2010:foldit}, software verification \cite{logas2014:xylem}, and nanomachine construction \cite{barone2015:nanocrafter}.
Additionally, HCGs have been used for crowdsourced data collection, such as photo acquisition for reconstruction of 3D buildings \cite{tuite2011:photocity}.
These are just a small handful of HCGs; recent taxonomies \cite{krause2011:hcg-survey, pe-than2013:hcg-typology} highlight the full breadth of human computation games and the tasks they have tackled.
Finally, while most HCGs have been built as standalone experiences, researchers have explored integration with mainstream digital games, such as Project Discovery \cite{peplow2016:eve-project-discovery}, which lets players of the game \emph{EVE Online} complete protein function recognition tasks in exchange for in-game rewards.

While human computation games have been shown to be an effective interface for solving crowdsourcing tasks, current design knowledge for developing these games remains limited.
Commonly-utilized guidelines for designing these games are often based on templates or anecdotal examples from successful games.
These include von Ahn and Dabbish's three game templates for classification and labeling tasks \cite{vonahn2008:gwap-design}, based on their early successes with HCGs, and the design of the gameplay mechanics in the protein-folding game \emph{FoldIt} \cite{cooper2010:sci-disc-design}.
However, it is not clear how these generalize to new kinds of tasks or changing player audiences.
Furthermore, there are no guidelines for ensuring game mechanics will guarantee both successful completion of the task and an engaging player experience.
Some researchers claim that HCGs should prefer game mechanics that map to the process of solving the underlying task \cite{jamieson2012:hcomp-games,tuite2014:gwap-problem}, while others argue that incorporating familiar or recognizable mechanics popular in digital games will keep players more engaged \cite{krause2010:ontogalaxy}.
This ongoing debate highlights the challenge of designing HCGs that optimize for both the human computation task and the player experience.

\subsection{Singleplayer and Multiplayer}
Researchers have studied the effects of singleplayer and multiplayer in the context of mainstream digital games, but not human computation games.
The gameplay of most HCGs is limited to singleplayer or networked multiplayer experiences; HCG players are generally not allowed to directly communicate during gameplay to avoid potential collusion \cite{vonahn2008:gwap-design}.

Research on co-location in games validates that players respond differently when playing with or against other human players, compared with singleplayer experiences or play against an artificial agent.
For example, Wehbe and Nacke \cite{wehbe2015:colocated-play} investigate the effect of co-location on players, finding that players demonstrated higher pleasure and perceived arousal in co-located multiplayer conditions than singleplayer conditions.
Similarly, Mandryk and Inkpen \cite{mandryk2004:co-located-play} report that players found co-located multiplayer play with a friend more engaging (i.e., more fun, less frustrating, less boring) than the same (singleplayer) experience against an artificial opponent.
Most relevant to HCGs because of their dual purpose, researchers have examined co-location in math games for education \cite{ke2007:gameplaying-for-maths, plass2013:impact-individual-collab-compete-math-games}.
Overall, results from these studies suggest that players demonstrate higher engagement in co-located multiplayer experiences when compared with singleplayer experiences.

\subsection{Collaboration and Competition}
Researchers have studied the effects of player collaboration (cooperation) in the context of multiplayer games, such as the work of Seif El-Nasr et al. identifying common patterns in cooperative play \cite{seifel-nasr2010:cooperative-games}.
Studies have looked at how collaboration and competition affect player experience metrics in motor performance games \cite{peng2012:motor-skills-coop-compete}, math games \cite{plass2013:impact-individual-collab-compete-math-games}, and co-located multiplayer games \cite{emmerich2013:loadstone}.

\begin{figure*}[tb]
 \centering
    \includegraphics[width=\textwidth]{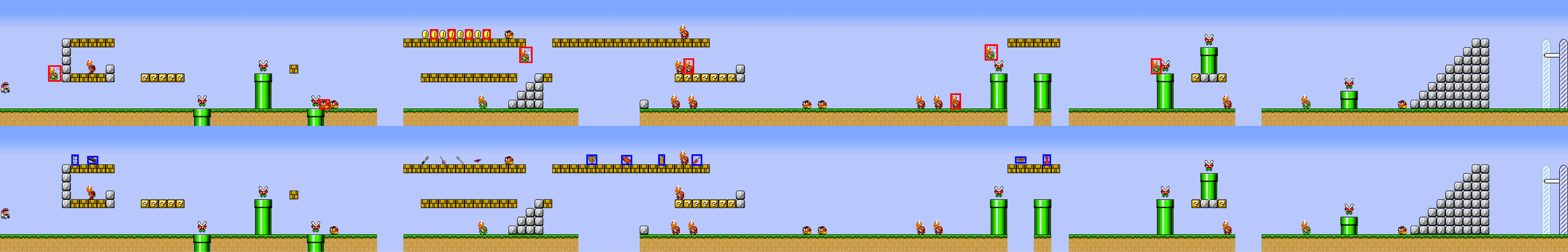}
    \caption{Comparison of the original level (top) and our edited level (bottom), with item/enemy additions in blue and deletions in red. Coins have been replaced with sprites of everyday items.}
    \label{fig:comparative}
\end{figure*}

In the context of human computation games, variations in collaboration and competition have been explored.
The game \emph{KissKissBan} \cite{ho2009:kiss-kiss-ban} modified the original collaborative form of the image-labeling \emph{ESP Game} \cite{vonahn2004:esp}, introducing competitive mechanics through an adversarial third player and yielding a greater variety of resulting labels.
Goh et al. \cite{goh2011:fight-unite} conducted a comparison of collaborative and competitive versions of the \emph{ESP Game} against a non-gamified control app, measuring the task results (number and quality of image labels) and player experience metrics such as appeal, challenge, and social interaction.
While the non-gamified app generated better task results, subjects preferred the game versions of the task, though no differences in player experience were found between the collaborative and competitive games.
Finally, Siu et al. \cite{siu2014:gwap} conducted a similar study comparing collaborative and competitive scoring systems in the context of a (simulated) networked multiplayer HCG.
They found no significant differences in completed task accuracy between collaborative and competitive scoring modes, but that players found the competitive mode more engaging.
Our results build on their investigations; while we are primarily interested in the effects of using singleplayer versus multiplayer mechanics, we also compare collaborative and competitive scoring systems in our multiplayer condition.

\section{The \textit{Gwario} Game}

We developed a human computation game based on the original \emph{Super Mario Bros.} (SMB), a game in which the player controls the titular Mario, attempting to clear a series of levels while jumping to avoid gaps and enemies. In addition to clearing levels, SMB contains a secondary objective: a numerical score, which can be improved by destroying enemies and collecting powerups and coins. Our HCG, \emph{Gwario} (a portmanteau of the term ``GWAP'' and ``Mario''), was built on our own observations of the original game and the code of the ``Infinite Mario'' game engine created by Markus Persson \cite{Anonymous:1976:POT}, within a novel game engine \cite{guzdial2016playable}. In order to better match the expectations of a modern audience, we made use of the equivalent visual components from a later game in the series, \emph{Super Mario World}, rather than using the original visual elements from SMB.

Traditionally, human computation games take the form of cooperative puzzle games \cite{vonahn2008:gwap-design} or gamified interfaces designed around the particular human computation task \cite{cooper2010:foldit, peplow2016:eve-project-discovery}.
HCGs are rarely classified in the genres of games designed purely for entertainment (with some exceptions such \emph{Ontogalaxy} \cite{carranza2012:ontogalaxy-vs-esp}, whose authors state the need to explore this issue).
We chose the original \emph{Super Mario Bros.} due to its preeminence among ``platformer'' games, both for its familiarity to players and its depth of study by the research community, as well as the availability of a secondary objective to use as the basis of gameplay mechanics for our human computation task.
HCGs are not typically designed to adopt mechanics from commercial games.
von Ahn and Dabbish \cite{vonahn2008:gwap-design} give a classical definition of HCGs as follows: ``[games] in which people, as a side effect of playing, perform tasks computers are unable to perform'' with a focus on ``useful output''.
The subsequent sections describe how \emph{Gwario} adheres to this definition.

\subsection{Adapting SMB to an HCG}
The original \emph{Super Mario Bros.} does not function as a human computation game. To adapt it to an HCG, we first selected a human computation task with a known solution: matching everyday items with a purchasing location (e.g., one might purchase ``breakfast cereal'' at the ``supermarket''). This task was used in prior HCG work comparing game mechanics \cite{siu2014:gwap} and we used the same gold-standard answer set. This classification or ``tagging" of items with purchasing locations is a problem for which we know the answer already, but it is useful for evaluating variations in mechanics because it allows us to measure accuracy of the results objectively (without attempting to simultaneously solve a novel human computation problem).

To incorporate the task into \emph{Gwario}, we altered \emph{Super Mario Bros.'s} existing secondary objective by replacing collectible coins with images of purchasable items as seen in Figure~\ref{fig:comparative}. Players of the game are assigned a purchasing location (e.g., ``supermarket''), and asked to collect only those items that could be bought at that location. Each playable section of the game (level) has twelve items, four of which correspond to one of three purchasing locations (``supermarket'', ``department store'', ``hardware store''). Players receive points towards their score for collecting any item, not only those that relate to their purchasing location, but we obfuscate this fact by displaying the score only at the end of each level. 

We made one additional significant departure from the original \emph{Super Mario Bros.}, adding a simultaneous two-player mode.
The original game had a multiplayer mode, but one in which two players switched back and forth with only one avatar on screen at a time. 
While this version of multiplayer is a significant departure from the original game, we based it upon later multiplayer implementations in the series such as that of the game \emph{New Super Mario Bros.}
In the two-player version, one player plays as ``Luigi'' with the other playing as Mario.
Within this multiplayer mode, we implemented two different scoring mechanics as two versions of the game: collaborative and competitive.
In the \emph{collaborative} version, both players' individual scores are combined at the end of each level.
In the \emph{competitive} version, we track both players' scores and display them individually at the end of each level.
In both multiplayer versions, each player is assigned a unique purchasing location; this allows us to better compare between singleplayer and multiplayer (as players are given the same location for both singleplayer and multiplayer).

\subsection{Game Levels}
For game levels, we wished to avoid audience familiarity with the original \emph{Super Mario Bros.} without compromising game design quality. 
Therefore, we chose four levels from \emph{Super Mario Bros.: The Lost Levels}, a Japan-exclusive sequel to the original SMB, which would be less familiar to an Western audience, and implemented them in our game engine.
As stated above, we changed the coins in each level to a unique purchasable item. In order to ensure task uniformity between the levels, we added items to each level that had less than twelve coins originally. We first removed items to break apart rows to avoid accidental collection. We then added items to locations similar to those that existed in the level initially, as demonstrated in Figure \ref{fig:comparative}. At the end of this process each level had an equal share of coins-changed-to-items from the three purchasing locations (``supermarket'', ``department store'', ``hardware store''). 

\emph{Super Mario Bros.: The Lost Levels} is noted for its intense difficulty.
Based on our own playthroughs of the levels, we anticipated this difficulty would negatively impact our study results.
To scale down the difficulty we made a series of initial changes: removing aerial enemies from the game and replacing especially difficult jumps.

We then ran a pilot study with ten subjects in five pairs to determine the appropriateness of our design changes. 
Based on the results of pilot study, we decreased the max jump distance required in the levels again, added powerups to the beginning of each level and decreased the density of enemy groups.
We present an example of one our final levels in comparison to its original in Figure~\ref{fig:comparative}.
Note that this level already had powerups (e..g, question mark blocks) at the beginning, so we did not need to add any.
\begin{figure*}[tb]
 \centering
    \includegraphics[width=\textwidth]{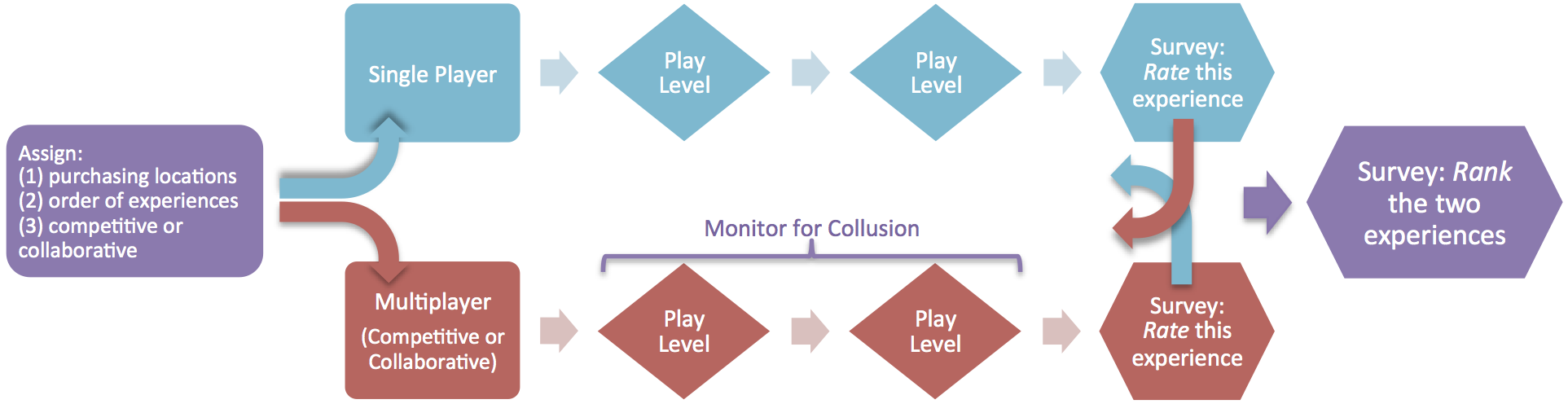}
    \caption{Flowchart of the methodology used in the experiment.}
    \label{fig:methodology}
\end{figure*}
\section{Methodology}
We ran a within-subjects study using the different versions of \emph{Gwario}.
Participants were able to sign up in pairs or could sign up individually to be paired by the experimenters.
We refer to these pairs of participants signed up together as self-selected or \emph{natural} pairings, as the participants were more likely to be acquainted prior to the experiment.
Pairs of participants who were scheduled together by the experimenters are referred to as \emph{artificial}, due to the fact that these participants were unlikely to have paired up to play a game together in a natural setting.

\subsection{Study}
The study consisted of two rounds of gameplay, followed by surveys after each round and a final survey following both rounds.
Each round of gameplay used a different game version, either singleplayer or multiplayer, and within multiplayer, either the collaborative or the competitive version. We visualize the flow of the study in Figure \ref{fig:methodology}.
Overall, we focused on emulating a casual play experience as much as possible. This guided our experimental design choices, such as tracking whether pairs of players were \emph{natural}, tracking collusion by hand instead of using microphones, and utilizing surveys over other measures.

Upon arrival for the study, pairs were randomly assigned to player either the singleplayer or multiplayer round of the game first.
For the multiplayer round, the pair was also randomly assigned either the collaborative or the competitive version.
As described previously, our game used four levels based on \emph{Super Mario Bros.: The Lost Levels}, adjusted slightly to compensate for the challenge of the original game.
To reduce the effects of ordering and difficulty, these levels were randomly assigned across the conditions upon arrival: two assigned for the singleplayer round and remaining two assigned for the multiplayer round. Each individual in the pair played the same two levels separately in the singleplayer round.
This random ordering of game conditions and levels, as well as the choice of collaborative or competitive multiplayer, was generated using a computer program to avoid bias.

For the singleplayer round, players were seated at separate computers and played through the game in isolation.
For the multiplayer round, players were seated at the same computer and presented with a single screen for gameplay (i.e., no split screen).
During the multiplayer round, players were told the victory condition --- either collaborative or competitive scoring --- and were also told that they could communicate if they desired.

During the multiplayer round, the individual running the study would tag whether or not collusion (i.e., discussion of the task) occurred along with the relevant quote or quotes. These quotes were later verified as instances of collusion by a second individual, with disagreements resolved via discussion.

There were no time limits imposed on play, however players were given three chances to restart upon death (i.e., ``lives'') for each level.
In the multiplayer version, if one player exhausted all of his or her lives, the remaining player was allowed to progress through the remainder of the level alone.
Finally, after each round, players were asked to answer several survey questions about their experience with that round.
After both rounds, players were given a final survey to establish demographic information and to compare their experience across both rounds.

\subsection{Evaluation Metrics}
To understand the efficacy of our game mechanic variations, we concentrated both on metrics to evaluate the \emph{player experience} and the \emph{task completion}, following a previously-suggested methodology for evaluating HCG design variations \cite{siu2014:gwap}.
Here, we outline what data and information was gathered.

To evaluate the \emph{player experience}, we report on the survey data, as well as gameplay events and reported collusion.
After each round (of singleplayer or multiplayer), players were asked Likert-style questions on a scale from 1-5 about their perceived fun/engagement, challenge, and frustration with the game.
After both rounds, players provided a comparative ranking between the two conditions (singleplayer and multiplayer) for perceived fun, challenge, frustration, and overall preference.
We also logged gameplay events such as when players died during the level, which players won/lost rounds, and end-of-level scores.
Additionally, we noted if participants communicated with each other during the multiplayer rounds.
A pair was noted to be \emph{colluding} if they discussed the human computation task at any point during the multiplayer round.

To evaluate the \emph{task completion}, we look at the results of our telemetry logging.
Task answers were logged and compared against our gold-standard answer set to determine correctness.
A player's accuracy at the given task was calculated as the percentage of their correctly-collected items over all correctly-collected items.
We also logged the number of tasks completed, as well as the times (in seconds) players took to answer tasks and complete the levels.
\section{Results}
We collected results from sixty-four individuals in thirty-two pairs over a two week period, advertising for subjects in two undergraduate computer science courses. 
Eighteen of the subjects identified as female and forty-six identified as male. 
Nearly all subjects were between the ages of eighteen and twenty-four, with three subjects older than twenty-five. Seventy-three percent of subjects reported that they played games regularly; all but ten had played a platformer game before. While this is a somewhat homogenous mixture of subjects, we contend that the majority of our subjects having played games makes this population a good stand-in for a typical HCG population. In addition, seventeen subjects reported having played a HCG before, which represents a significant number of experienced HCG players.

Our study focused on co-located pairs of subjects.
Fifteen of the thirty-two pairs were \emph{natural}, meaning that both individuals signed up to take the study together purposely.
The remaining seventeen pairs were \emph{artificial}, consisting of subjects who signed up to take the study without a partner and then were randomly assigned an available partner and time slot.
While we were concerned that differences (i.e., prior acquaintanceship with a partner) might have impacted our results, we found no significant differences between \emph{natural} and \emph{artificial} pairs of subjects across any of our subjective (player experience) or objective (task) metrics described below.

\subsection{Subjective Metrics}
\begin{figure}[tb]
 \centering
    \includegraphics[width=\linewidth]{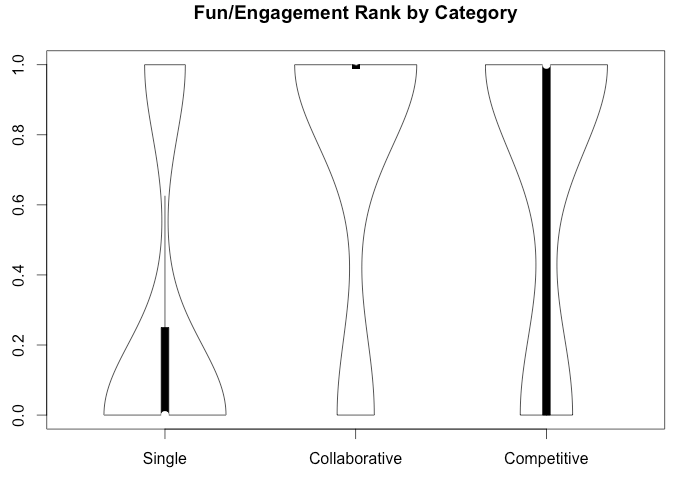}
    \caption{Violin plot of fun/engagement \emph{ranking} across the gameplay variations from 1.0 (most) to 0.0 (least) fun/engaging. The diameter across the length of the violin indicates the number of results of that value. The dark bar in each violin runs between the first and third quartiles.}
    \label{fig:engagementRank}
\end{figure}

We collected responses on subjective experience in terms of fun (engagement), frustration, challenge, and overall experience with five-point Likert ratings and rankings between singleplayer and multiplayer. We were unable to find any significant differences comparing Likert ratings (using the paired Wilcoxon Mann-Whitney U test) except in one case. Specifically we found that subjects in the competitive condition rated the multiplayer round significantly more challenging than the singleplayer round ($p < 0.01$). The ranking data was much more discriminatory. Subjects across both conditions ranked multiplayer as being more fun/engaging as seen in Figure~\ref{fig:engagementRank} and overall preferring it to singleplayer according to the paired Wilcoxon Mann-Whitney U test($p < 0.05$). In addition in the competitive condition, subjects rated multiplayer as more challenging than singleplayer, which confirms the challenge rating result ($p <0.01$). 

We expected the competitive mode to be more challenging, as subjects have to compete against the game and each other. These results further demonstrate that instructions and a simple alteration in presenting scores is sufficient to invoke competitive behavior. Despite these subjective experience reports, no significant difference exists in the number of player deaths between competitive multiplayer and singleplayer.

\begin{table}[tbh]
\begin{tabular}{ l || p{0.2\linewidth} | p{0.2\linewidth} | p{0.2\linewidth}}
 & \small Singleplayer & \small Collaborative Multiplayer & \small Competitive Multiplayer \\
\hline
Accuracy Avg. &  82\% & 86\% & 77\%\\
Time(s) Avg. &  212 & 612 & 505\\
Tasks Avg.  &  22 & 16 & 14\\
Deaths Avg.  &  4.9 & 4.7 & 5.1\\
\end{tabular}
\caption{Summary of Objective Results}
\label{table:descriptiveElements}
\end{table}
\begin{figure}[tb]
 \centering
    \includegraphics[width=\linewidth]{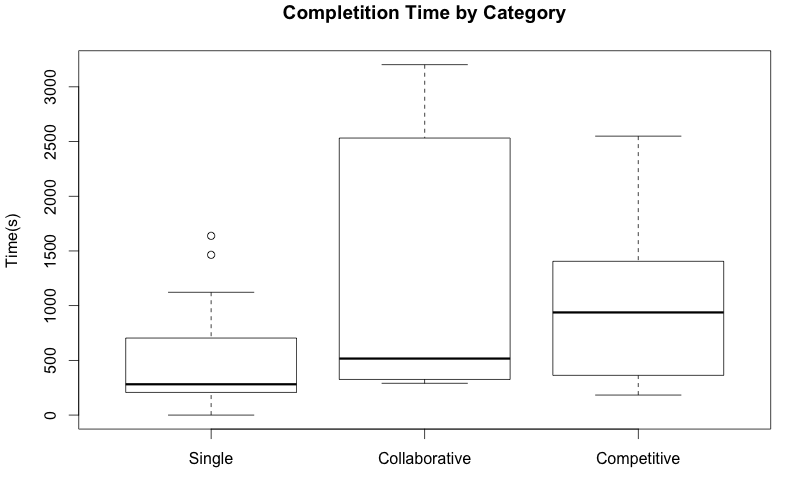}
    \caption{Plot of time to level completion across variations}
    \label{fig:timeMetric}
\end{figure}
\subsection{Objective Metrics}
We tracked two objective metrics to measure subject performance on the human computation game: the time it took a subject or pair of subjects to complete a level (in victory or defeat) and the per-task accuracy of each subject (the ratio of correct task assignments to total tasks attempted). On average across conditions of the study the subjects had a 81.7\% per-task accuracy. In addition subjects completed 4.4 tasks per attempt, meaning each subject collected an average of 4.4 items per life, when 4 of the 12 items actually corresponded to a given subject's purchasing location. Taken together this suggests that the subjects were able to perform fairly well at the task, despite having to also simultaneously play through a Mario level. 

We summarize the objective metrics by condition in Table~\ref{table:descriptiveElements}. From these average values we can identify some major distinctions between the two multiplayer conditions relative to a singleplayer experience. In terms of accuracy, we can see that collaborative multiplayer has the highest, though there is no significant difference between any of the accuracy distributions. However, there is a significant difference between the collaborative and competitive conditions, with the collaborative condition having a significantly greater change in accuracy between the singleplayer and multiplayer modes according to the Wilcoxon Signed Rank test ($p <0.05$). In other words, subjects were more accurate in the collaborative multiplayer mode than in the singleplayer mode, and less accurate in competitive multiplayer.

\begin{sloppypar}
The average times in Table~\ref{table:descriptiveElements} suggest that collaborative playthroughs took much longer than either singleplayer or competitive playthroughs, but this is not true for most cases. Figure~\ref{fig:timeMetric} demonstrates that collaborative playthroughs had a far wider distribution of times and a higher upper bound, but that the median collaborative time is below the median competitive time. The table also shows that players accomplished the most tasks on average in the singleplayer mode, second to collaborative and then competitive multiplayer. Lastly we include the average number of deaths per subject across both levels. Despite subjects consistently rating and ranking competitive as more challenging than singleplayer, we found no significant difference in the number of deaths between the two conditions.
\end{sloppypar}

We sought an explanation for the variance in accuracy between the competitive and collaborative conditions. We ran an ANOVA with permutations, due to the non-ordinal data, looking to predict multiplayer accuracy across both conditions. From this test we found that collusion, whether or not the subjects spoke about the task, was the only significant predictor of accuracy across multiplayer condition, whether the pair was \emph{natural}, and all demographic information ($r = 12.24; p < 0.01$). This may also suggest why the collaborative condition had the greatest variance in time, given the time it took subjects to discuss the tasks with one another.
\section{Expert Opinions}
To better contextualize our findings in the space of human computation game design, we pursued expert opinions on the mechanical variations we tested in \emph{Gwario}.
We wanted to ensure that our experimental design spoke to the intuitions and uncertainties of current HCG researchers and developers.

We identified experts who were responsible for developing nine recent human computation games (including those described in the related work) and sent them a survey asking for their opinions on the mechanics we tested in our study.
An expert was considered to be someone who had developed an HCG and (with one exception) had peer-reviewed publications on the findings or design of the game.
None of the experts we queried were made aware of the results of the \emph{Gwario} study prior to taking the survey.

The survey covered four topics.
First, we asked about three different variations in HCG mechanics: singleplayer versus co-located multiplayer, collaborative versus competitive scoring, and prohibiting versus permitting direct player communication.
For each variation, we asked experts how they thought a hypothetical game employing the second condition would compare to the first along two metrics: task accuracy and player engagement.
For each metric, experts chose from three multiple choice answers: ``increased (accuracy/engagement)'', ``no difference in (accuracy/engagement)'', and ``decreased (accuracy/engagement)''.
Experts were also asked to provide short-answer descriptions about their choices.
Lastly, experts were asked a final question---if mechanics from successful digital games should be incorporated into HCGs (``yes/no/maybe'')---and to provide a short-answer description about why.

Three experts responded to our survey.
While we acknowledge that these results are limited, we find their information to be a valuable insight into current perceptions of HCG design.
The results of these questions are shown in Table~\ref{table:experts}.

\begin{table}[tbh]
\begin{tabular}{ l || p{0.25\linewidth} | p{0.25\linewidth} | p{0.25\linewidth} }
 & \small \textbf{Multiplayer} would be & \small \textbf{Competition} would be & \small \textbf{Direct communication} would be \\
\hline
Expert 1 & \small no more accurate & \small more accurate & \small more accurate\\
& \small more engaging & \small no more engaging & \small more engaging\\
\midrule
Expert 2  & \small more accurate & \small no more accurate &\small \small more accurate\\
& \small more engaging & \small more engaging & \small more engaging\\
\midrule
Expert 3  & \small more accurate & \small less accurate & \small more accurate\\
& \small more engaging & \small no more engaging & \small more engaging\\
\hline
 & \small than \textbf{singleplayer} & \small than \textbf{collaboration} & \small than \textbf{no communication} \\
\end{tabular}
\caption{Results of the expert survey asking about expected accuracy and engagement for variations in HCG mechanics.}
\label{table:experts}
\end{table}

When it came to singleplayer versus co-located multiplayer, experts agreed that co-located multiplayer would increase (or have no difference in) accuracy and player engagement.
This suggests that co-located multiplayer is perceived as more effective and beneficial than singleplayer.
Expert~3 compared the benefits to pair-programming, but expressed that ``local, as opposed to remote, co-operative games are harder to coordinate''.

When looking at collaborative versus competitive scoring, experts did not agree on how it would affect task accuracy.
Expert~1 noted that ``competition could increase accuracy for certain tasks because it gives players a way to measure themselves and their contributions''.
By contrast, Expert~3 noted that ``competitive players will find ways to win which do not advance scientific goals if they are more expedient'', suggesting that adding competition would negatively affect non-competitive players while competitive players would remain unaffected.
Regarding player engagement, experts were also mixed, but agreed that competition would not decrease player engagement.
Expert~2 cited examples of HCGs where competition was shown to have positive benefits on engagement.

When looking at prohibiting versus permitting direct player communication, experts agreed that direct player communication would lead to both increased task accuracy and player engagement.
Expert~1 did caution that allowing direct communication would not work ``if the mechanics are directly related to players independently coming up with ideas (e.g., \emph{ESP Game})'', but noted that not all HCGs follow the same format.

Finally, all experts agreed that mechanics from successful digital games should possibly be incorporated into human computation games, with Expert~1 stating ``maybe'' and Experts~2~\&~3 stating ``yes''.
When asked why, experts focused on player familiarity with game mechanics, but noted possible concerns with incorporating these mechanics into HCGs.
In particular, Expert~1 expressed concern that mechanics that did not compliment the task might compromise the task. Expert~3 remarked that mapping mechanics from successful games (where players have different motivations for play) to HCGs remains an open question.
\section{Discussion}
In this section we summarize the major results of our human subject study, the expert opinion survey, and compare our results with those of similar prior work. Given our focus on design, we organize the discussion section into a set of design implications for \emph{Gwario}, tied to its dual goals of creating an engaging player experience and solving the human computation task. 

\subsection{Transforming ``Collect-a-thons'' into HCGs}
We define a ``collect-a-thon'' as any game where exploration for collectible items (e.g. coins, rings, notes) is a major gameplay \emph{mechanic} (e.g., classic game series such as Mario, Sonic, and Banjo Kazooie). In Gwario, we convert the collectible elements into items we desire to categorize or classify. Thus, the player's choice to collect an item maps to the process of answering human computation task, making such mechanics most appropriate for classification or categorization tasks. Outside of \emph{Gwario}, \emph{OnToGalaxy} \cite{krause2010:ontogalaxy} is an HCG that implemented a similar game design, as its design is that of an archetypal ``space shooter'' transformed into a HCG by altering collectable objects into task answers. Notably in this example the categorization tasks are much fuzzier (e.g. collect [items] labeled with phrase ``touchable objects").

Experts suggest that incorporating mechanics from successful digital games should be considered for HCGs, but caution that the mechanics should be appropriate for the corresponding human computation task. Our implementation of this design maps \emph{collection} mechanics to \emph{categorization}, and appears to successfully marry the twin design goals of \emph{player experience} and \emph{task completion}. Across conditions we found an average accuracy of more then eighty percent and a consistent median Likert rating of ``4'' for fun/engagement on a five-point Likert scale. This is strong evidence that this design retains much of the fun and familiarity of the base game. However, from the results of our pilot study we would expect the impact of using this design to vary wildly depending on how challenging the base game is, notably whether or not players are given the appropriate decision-making time to consider categorization choices.

\subsection{Affording communication in real-time multiplayer HCGs}
We now highlight the direct communication between players during a real-time multiplayer HCG. This could be in the form of in-person conversation, audio or video chat, a text-only interface, or some discretized set of allowable messages. This game design goes against typical HCG design considerations, which suggest that collusion between players could hurt task accuracy \cite{vonahn2008:gwap-design, downs2010:mturk-gaming-the-system}. Therefore to the best of our knowledge, \emph{Gwario} is the only current example of this design, wherein co-located direct communication is permitted \emph{during the process of completing the task}. However, asynchronous and indirect communication have been explored in HCGs. For example, \emph{FoldIt} \cite{cooper2010:sci-disc-design} permits asynchronous communication through player forums and solution-sharing, representing a related design. 

Direct communication has been shown to benefit games from other dual-purpose domains, such as educational math games \cite{plass2013:impact-individual-collab-compete-math-games} where side-by-side student play increases learning motivation and engagement. In addition, outside of games entirely this design manifests in the CS Education practice of \emph{pair-programming}, in which two individual students solve a programming problem at the same time and has shown impressive positive results \cite{salleh2011empirical}. This analogy was cited in our expert opinions (Expert~3).

Our expert surveys suggest that direct-communication between players is perceived to have benefits for both player engagement and task accuracy. We found direct evidence of player communication's impact on task accuracy, as player collusion was a significant predictor of increased accuracy. While we did not find similar evidence linking communication and engagement, the significantly higher ranking of multiplayer fun/engagement suggests a potential connection. These findings from both our HCG and expert feedback question the assumptions made by prior HCG research that suggest players may try to ``game'' the game to gain more rewards (e.g. collecting all coins in \emph{Gwario}, regardless of correctness to improve their end score).
Note that even if data-tainting collusion did occur in a game instantiating this design, one could cross-validate the results across many dyads to effectively neutralize it.

Incorporating mechanics that permit direct communication may be sensitive to the particular task and gameplay mechanics, and we believe further verification of this design is needed. It may be that points are less a motivator than the constraints of the HCG, or the primary gameplay. In addition, while the most predictive factor of accuracy was our boolean measure of collusion, the most predictive factor of collusion was the multiplayer condition. In other words, players instructed to collaborate were much more likely to positively collude. Outside of HCGs, this design could prove useful in other serious game applications, as similar effects have been shown in educational games.

\subsection{Synchronous \emph{Competitive} Multiplayer HCGs}
We highlight the mechanics facilitating competitive multiplayer in an HCG. In particular we specify \emph{synchronous} competitive multiplayer, where two player make decisions simultaneously in real-time in an effort to outperform one another. In this context, outperforming another player refers to achieving a higher score (a typical HCG mechanic) than that of their opponent. Many HCGs have competitive elements in the form of leaderboards, but synchronous competition is less common. We highlight two examples. In \emph{KissKissBan} \cite{ho2009:kiss-kiss-ban}, a third player synchronously competes with two other players who are working together collaboratively; however, all three players are contributing answers to the underlying image-labeling task. In both Goh et al.'s \emph{ESP Games} \cite{goh2011:fight-unite} and \emph{Cabbage Quest} \cite{siu2014:gwap}, players either compete or collaborate to tag items as quickly as possible. These games have a similar experimental setups to our work. However, \emph{Cabbage Quest} diverges from \emph{Gwario} as players do not compete in person, but against an artificial player they are led to believe is another human.

We found strong results that competitive multiplayer is viewed as significantly more challenging and more fun/engaging than singleplayer in \emph{Gwario}, which is in line with the results from \emph{KissKissBan} and \emph{Cabbage Quest}. This suggests that adding competitive elements could promote a more positive \emph{player experience}, especially with a simple (and mundane) task (as in \emph{Cabbage Quest}). Meanwhile, Goh et al. found no difference in player engagement between competitive and collaborative versions of the \emph{ESP Game}, suggesting that competitive gameplay elements were no worse than their collaborative counterparts. Altogether, this mirrors expert opinions, which suggest that competition will not negatively impact player engagement.

The effects of competitive game mechanics on \emph{task completion} are potentially negative. In particular, we highlight that one danger with competitive mechanics in HCGs is a potential negative impact on accuracy. This is reinforced with the differing expert opinions about how such mechanics would affect accuracy. One expert-suggested benefit of competition is the potential feedback for players to measure themselves and their contributions, but one expert-suggested detriment would be that this would encourage expedient (but not necessarily correct) solutions. In our study, we found strong evidence that competitive players were significantly less accurate than collaborative players, however both Goh et al.'s \emph{ESP Games} and \emph{Cabbage Quest} found no such significant difference. This could be due to a number of differences between \emph{Gwario} and these two games. In Goh et al.'s competitive \emph{ESP Game} and \emph{Cabbage Quest}, players were not co-located, could not communicate, and had fewer mechanics available to antagonize the other player. In competitive \emph{Gwario}, players could antagonize (impede) each other by stealing power-ups from one another, attempting to hurt/kill each other with shells, and jumping around to distract their opponent. While some players seemed enjoy these affordances, the existence of these mechanics may have led to the poorer accuracy compared to that observed in \emph{Cabbage Quest}. In addition, we found that competitive multiplayer play took significantly more time than singleplayer play, which may be a deterrent towards implementing competitive mechanics. However, if rate of result acquisition is not a design concern and the accuracy issue can be avoided via cross-validation, this competitive mechanics seems may prove to increase player engagement and a sense of challenge.
\subsection{Synchronous \emph{Collaborative} Multiplayer HCGs}
We identify the mechanics of collaborative, synchronous multiplayer in an HCG. Collaborative play in an HCG is facilitated through the inclusion of mechanics that require two or more players to work together to improve the same in-game reward (i.e., score) or end result. Nearly all synchronous, multiplayer human computation games, dating back to the original \emph{ESP Game} \cite{vonahn2004:esp}, reward players for collaboration. This paradigm maps to the structure of the human computation process, in which verification of the result may be accomplished through aggregated agreement. In HCGs, this manifests as mechanics which tie together the verification of an HCG task and scoring, as players have to agree on a tag before receiving their reward (typically points). The collaborative version of \emph{Gwario} is similar, but differs slightly from these historical examples, as both players have separate categorizing tasks and a discrete pool of possible categories (ensuring that omitting an object from a category is verification that it may belong to another).

We implemented this design in \emph{Gwario} and our results match previous expectations given the historical use of collaboration in HCGs. Players found collaborative multiplayer more fun/engaging and overall preferred it to singleplayer. In comparison to competitive mechanics, prior work has suggested that competitive mechanics may be more engaging, but that certain aspects of the player experience may be higher for collaborative play (e.g., player empathy in the collaborative version of \emph{Loadstone} \cite{emmerich2013:loadstone}). Meanwhile, our experts were neutral (with one exception) on the idea that collaborative multiplayer was more engaging that competitive multiplayer. We cannot directly compare player fun/engagement in collaborative and competitive play, as individual players only played one variation. Players were slightly more likely to rank collaborative as most fun/engaging in comparison to the competitive ranking, but this was not significant. Further work is required to tease out these potential variations.

When combining these collaborative mechanics and affording communication during real-time multiplayer HCGs, we found a significant increase in accuracy in comparison to the competitive variation and the highest average accuracy overall. These results differ from Goh et al.'s \emph{ESP Games} and \emph{Cabbage Quest}, which found no significant difference in accuracy between their competitive and collaborative variations. However, the difference here is likely due to our pairing of design features, perhaps in conjunction with our co-located implementation of these mechanics. Notably, we find strong evidence that our variation of this design, in giving individuals two distinct tagging tasks as opposed to working towards agreement on a single task, has proven successful in terms of accuracy, time, and players' self reported fun/engagement. This suggests the potential for such collaborative mechanics to reduce the size of the player base needed to solve an HCG.

\section{Limitations}
We make note of limitations and potential for future work implicit in our results. Foremost, our results come from variations of a single game, \emph{Gwario}. This focus allows us to simulate the impact of social and mechanical changes in development of a novel HCG, but makes arguments of generality difficult. We lessen this potential impact by gathering expert opinions about our specific game mechanic variations and identifying how our results compare to prior HCGs, but further validation in novel HCGs (especially those that adapt mechanics from successful digital games) is still needed. 

We found a set of significant tradeoffs between competitive and collaborative play in \emph{Gwario}. However these results are only directly relevant to co-located competitive and collaborative play. Whether these results hold true for networked play is unknown, though we can speculate given that singleplayer \emph{Gwario} is functionally similar to a na\"ive, networked implementation, as both omit any means of direct communication between players (and traditionally, networked HCGs do not permit direct player communication).
Similarly to other studies of co-location \cite{mandryk2004:co-located-play, wehbe2015:colocated-play}, our study was conducted in a lab environment.
While we took steps to mitigate the encumbrances of a lab environment and provide a non-intrusive, comfortable space for play, we acknowledge this may not have been the most natural space for gameplay.
A future investigation comparing a networked multiplayer version of \emph{Gwario} to our co-located results would both help to verify our speculations and address the limitations of conducting the study in a lab setting.
\section{Conclusions}
In this paper, we describe a study of gameplay mechanics in human computation games.
We developed an HCG, \emph{Gwario}, by adapting the mechanics of \emph{Super Mario Bros.} and used it to test variations in singleplayer and multiplayer, along with collaborative and competitive scoring.
In doing so, we explore how collection mechanics from successful digital games (e.g., SMB) could be adapted to HCGs.

We conducted a study using \emph{Gwario} and observed the effects of these variations on aspects of the \emph{player experience} and the \emph{task completion}.
We also surveyed HCG experts to better understand their perceptions and insight of current HCG design practices.
Our results for the \emph{player experience} show that competitive multiplayer was seen as significantly more challenging than singleplayer or collaborative multiplayer, while multiplayer was considered more engaging than singleplayer.
For \emph{task completion}, players in collaborative multiplayer were the most accurate at the task, while those in competitive multiplayer were the least accurate.
Finally, we found that collusion (i.e., direct communication) was a significant predictor of high task accuracy.
We discuss these results in the context of prior HCGs and collected expert opinions.
We draw attention to the benefits and tradeoffs of collaboration and competition, as well as potential contradictions of existing HCG design hypotheses such as the allowance of direct communication.

Human computation games have shown great promise as an interface for solving human computation tasks, but generalized design knowledge to empower novice HCG designers has been lacking.
As a result, these games remain unexplored and their full potential is undetermined.
Our work intends to help broaden the knowledge base of HCG design, making development of these games more accessible to researchers and task providers.
In doing so, we hope to expand the use and quality of HCGs in ways that benefit both players and the problems they solve through play.

\section*{Acknowledgements}
We thank members of the Entertainment Intelligence Lab for providing feedback on the study and the game.
We also thank our players for participating in this study, and our surveyed HCG experts for their time and contributions.

This material is based upon work supported by the National Science Foundation under Grant No. 1525967. Any opinions, findings, and conclusions or recommendations expressed in this material are those of the author(s) and do not necessarily reflect the views of the National Science Foundation.

\bibliographystyle{ACM-Reference-Format}
\bibliography{proceedings-fdg}

\end{document}